\documentclass{sig-alternate-05-2015}
\usepackage{pdfpages}

\usepackage{graphicx}
\usepackage[flushleft]{threeparttable}
\usepackage[FIGTOPCAP]{subfigure}
\usepackage{xspace}
\usepackage{multirow}
\usepackage{booktabs}
\usepackage{pifont}
\usepackage{subfig}

\usepackage{tikz}
\usetikzlibrary{decorations.markings}
\usetikzlibrary{shapes.geometric }
\usetikzlibrary{arrows}

\usepackage{ifthen,amssymb,color}
\newboolean{showcomments}
\setboolean{showcomments}{false}
\ifthenelse{\boolean{showcomments}}
  {\newcommand{\nb}[2]{
  \fbox{\bfseries\sffamily\scriptsize#1}
    {\sf\small$\blacktriangleright$\textit{\textcolor{red}{#2}}$\blacktriangleleft$}
   }
  }
  {\newcommand{\nb}[2]{}
   
  }

\begin{document}

\newcommand{\tool}{\emph{AnFlo}\xspace}
\newcommand{\reference}{trusted\xspace}
\newcommand{\Reference}{Trusted\xspace}
\newcommand{\numFlows}{25\xspace}
\newcommand{\ACCU}{accurate\xspace}
\newcommand{\ACCUN}{accuracy\xspace}

\newcommand{\GAPP}{\emph{benign apps}\xspace}
\newcommand{\VPATH}{candidate vulnerable path\xspace}
\newcommand{\VPATHS}{candidate vulnerable paths\xspace}
\newcommand{\AUT}{AUA\xspace}
\newcommand{\AUTs}{AUAs\xspace}
\newcommand{\PREV}{\emph{PREV}\xspace}
\newcommand{\FLOWD}{\emph{FlowDroid}\xspace}
\newcommand{\TAINTD}{\emph{TaintDroid}\xspace}
\newcommand{\numAUTTotal}{596\xspace}
\newcommand{\numAUTWithFlows}{91\xspace}
\newcommand{\numAUTWithRelevantFlows}{76\xspace}
\newcommand{\normal}{normal\xspace}
\newcommand{\Normal}{Normal\xspace}

\newcommand{\numApps}{11,796\xspace}
\newcommand{\avgnumAppsPerCat}{500\xspace}

\title{
AnFlo: Detecting Anomalous Sensitive Information Flows in Android Apps}

\numberofauthors{2}

\author{
\alignauthor
Biniam Fisseha Demissie, Mariano Ceccato\\
       \affaddr{Fondazione Bruno Kessler}\\
       \affaddr{Trento, Italy}\\
       \email{\{demissie,ceccato\}@fbk.eu}
       \alignauthor
Lwin Khin Shar\\
       \affaddr{School of Computer Science and Engineering}\\
       \affaddr{Nanyang Technological University, Singapore}\\
       \email{lkshar@ntu.edu.sg}
}

\CopyrightYear{2018} 
\setcopyright{acmcopyright} 
\conferenceinfo{MOBILESoft '18,}{May 27--28, 2018, Gothenburg, Sweden}
\isbn{978-1-4503-5712-8/18/05}\acmPrice{\$15.00}
\doi{https://doi.org/10.1145/3197231.3197238}

\maketitle

\begin{abstract}

Smartphone apps usually have access to sensitive user data such as contacts, geo-location, and account credentials and they might share such data to external entities through the Internet or with other apps. 
Confidentiality of user data could be breached if there are anomalies in the way sensitive data is handled by an app which is vulnerable or malicious. Existing approaches that detect anomalous sensitive data flows have limitations in terms of \ACCUN because the definition of \emph{anomalous flows} may differ for different apps with different functionalities; it is \normal for ``Health'' apps to share heart rate information through the Internet but is anomalous for ``Travel'' apps.

In this paper, we propose a novel approach to detect anomalous sensitive data flows in Android apps, with improved \ACCUN. To achieve this objective, we first group \emph{\reference} apps according to the topics inferred from their functional descriptions. We then learn sensitive information flows with respect to each group of \reference apps. 
For a given app under analysis, anomalies are identified by comparing sensitive information flows in the app against those flows learned from \reference apps grouped under the same topic. 
In the evaluation, information flow is  learned from \numApps \reference apps. We then checked for anomalies in \numAUTTotal new (benign) apps and identified 2 previously-unknown vulnerable apps related to anomalous flows. We also analyzed 18 malware apps and found anomalies in 6 of them.

  \end{abstract}

\section{Introduction}
\label{sec:intro}

Android applications (apps) are often granted access to users' privacy- and security-sensitive information such as GPS position, phone contacts, camera, microphone, training log, and heart rate. Apps need such sensitive data to implement their functionalities and provide rich user experiences. For instance, accurate GPS position is needed to navigate users to their destinations, phone contact is needed to implement messaging and chat functionalities, and heart rate frequency is important to accurately monitor training improvements.

Often, to provide services, apps may also need to exchange data with other apps in the same smartphone or externally with a remote server. For instance, a camera app may share a picture with a multimedia messaging app for sending it to a friend. The messaging app, in turn, may send the full contacts list from the phone directory to a remote server in order to identify which contacts are registered to the messaging service so that they can be shown as possible destinations. 

As such, sensitive information may \emph{legitimately} be propagated via message exchanges among apps or to remote servers. 
On the other hand, sensitive information might be exposed unintentionally by  defective/vulnerable apps or intentionally by malicious apps (malware), which threatens the security and privacy of end users. 
Existing literature on information leak in smartphone apps tend to overlook the difference between legitimate data flows and illegitimate ones. Whenever information flow from a sensitive source to a sensitive sink is detected, either statically~\cite{amandroid}, \cite{epicc}, \cite{ ic3,iccta,flowdroid}, \cite{jitana}, \cite{Mann2012privacyLeaks}, \cite{didfail} or dynamically~\cite{Enck2010taintDroid}, it is reported as potentially problematic. 

In this paper, we address the problem of detecting anomalous information flows with improved \ACCUN by classifying cases of information flows as either {\em \normal} or {\em anomalous} according to a reference information flow model. More specifically, we build a model of sensitive information flows based on the following features:
\begin{itemize}
\item Data source: the provenance of the sensitive data that is being propagated; 
\item Data sink: the destination where the data is flowing to; and
\item App topic: the declared functionalities of the app according to its description.
\end{itemize}
\emph{Data source} and \emph{data sink} features are used to reflect information flows from sensitive sources to sinks and summarize how sensitive data is handled by an app. However, these features are not expressive enough to build an \ACCU model. In fact, distinct apps might have very different functionalities. What is considered legitimate of a particular set of apps (e.g., sharing contacts for a messaging app) can be considered a malicious behavior for other apps (e.g., a piece of malware that steals contacts, to be later used by spammers). An \ACCU model should also take into consideration the main functionalities that is declared by an app (in our case the {\em App topic}). One should classify an app as anomalous only when it exhibits sensitive information flows that are not consistent with its declared functionalities. This characteristic, which makes an app anomalous, is captured by the \emph{App topic} feature.

In summary, our approach focuses on detecting apps that are anomalous in terms of information flows compared to other apps with similar functionalities. Such an approach would be useful for 
 various stakeholders. For example, market owners (e.g., Google) can focus on performing more complex and expensive security analysis only on those cases that are reported as anomalous, before publishing them. 
If such information is available to end users, they could also make informed decision of whether or not to install the anomalous app. For example, when the user installs an app, a warning stating that this particular app sends contact information through the Internet differently from other apps with similar functionalities (as demonstrated in the tool website).
In the context of BYOD (bring your own device) where employees use their own device to connect to the secure corporate network, a security analyst might benefit from this approach to emphasis manual analysis on those anomalous flows that might compromise the confidentiality of corporate data stored in the devices.

The specific contributions of this paper are:
\begin{itemize}
	\item An automated, fast approach for detecting anomalous flows of sensitive information in Android apps through a seamless combination of static analysis, natural language processing, model inference, and classification techniques;
	\item The implementation of the proposed approach in a tool called \tool which is publicly available\footnote{Tool and dataset available at http://selab.fbk.eu/anflo/};  and
	\item An extensive empirical evaluation of our approach based on \numAUTTotal subject apps, which assesses the \ACCUN and runtime performance of anomalous information flow detection. We detected 2  previous-unknown vulnerable apps related to anomalous flows. We also analyzed 18 malware apps and found anomalies in 6 of them.
\end{itemize}

The rest of the paper is organized as follows. 
Section~\ref{sec:motivation} motivates this work. 
Section~\ref{sec:related} compares our work with literature.
Section~\ref{sec:model} first gives an overview of our approach and then explains the steps in details. 
Section~\ref{sec:results} evaluates our approach.
Section~\ref{sec:conclusion} concludes the paper.
 
\section{Motivation}
\label{sec:motivation}

To implement their services, apps may access sensitive data. It is important that application code handling such data follows secure coding guidelines to protect user privacy and security. However, fast time-to-market pressure often pushes developers to implement data handling code quickly without considering security implications and release apps without proper testing. As a result, apps might contain \emph{defects} that leak sensitive data unintentionally.
They may also contain \emph{security vulnerabilities} such as permission re-delegation vulnerabilities~\cite{Porter2011re-delegation}, which could be exploited by malicious apps installed on the same device to steal sensitive data. 
Sensitive data could also be intentionally misused by \emph{malicious} apps.
Malicious apps such as malware and spyware often implement hidden functionalities not declared in their functional descriptions. For example, a malicious app may declare only entertainment features (e.g., games) in its description, but it steals user data or subscribes to paid services without the knowledge and consent of the user.

Defective, vulnerable, and malicious apps all share the same pattern, i.e., they (either intentionally or unintentionally) deal with sensitive data in an anomalous way, i.e., they behave differently in terms of dealing with sensitive data compared to other apps that state similar functionalities.  
Therefore, novel approaches should focus on detecting anomalies in sensitive data flows, caused by mismatches between expected flows (observed in benign and correct apps) and actual data flows observed in the app under analysis. However, the comparison should be only against {\em similar} apps that offer similar functionalities. For instance, messaging apps are expected to read information from phone contact list but they are not expected to use GPS position. 

These observations motivate our proposed approach.

\section{Related Work}
\label{sec:related}

The approaches proposed in Mudflow~\cite{Mudflow:ICSE15} and Chabada~\cite{gorla2014checking} are closely related to ours. Mudflow~\cite{Mudflow:ICSE15} is a tool for malware detection based on sensitive information flow. Similar to our approach, they rely on static taint analysis to detect flows of sensitive data towards potential leaks. Then, these flows are used to train a $\nu-$SVM one-class classifier and later classify new apps. While we also use static analysis, the main difference is that we consider the dominant topic inferred from app description as an important feature for the classification. Our empirical evaluation shows that dominant topics are fundamental to achieve a higher \ACCUN in anomaly detection. Moreover, our approach not only focuses on detecting malware, but also focuses on vulnerable and defective apps. Lastly, while Mudflow applies intra-component static analysis, we use inter-component analysis for covering flows across components.

Chabada~\cite{gorla2014checking} is a tool to find apps whose descriptions differ from their implementations. While we apply similar techniques in terms of natural language processing of apps descriptions, the goals differ. The goal of their approach is to find anomalous apps among the apps in the wild based on the inconsistencies between the advertised apps descriptions and their actual behaviors. By contrast, our approach specifically targets at  identifying anomalies in the flow of sensitive information in the app code. More specifically, Chabada only identifies calls to sensitive APIs to characterize benign and anomalous apps. By contrast, we consider data flows  from sensitive data sources to sensitive sinks.

Information leak in mobile apps is a widespread security problem. Many approaches that deal with this security problem are related to ours. Information flow in mobile apps is analysed either statically~\cite{amandroid}, \cite{epicc}, \cite{ic3}, \cite{jitana}, \cite{Mann2012privacyLeaks}, \cite{iccta}, \cite{didfail} or dynamically~\cite{Enck2010taintDroid}, to detect disclosure of sensible information. Tainted sources are system calls that access private data (e.g., global position, contacts entries), while sinks are all the possible ways that make data leave the system (e.g., network transmissions). An issue is detected when privileged information could potentially leave the app through one of the sinks. In the following, we discuss some of these approaches and then explain the major differences.

Amandroid~\cite{amandroid} has been proposed to detect privacy data leaks due to inter-component communication (ICC) in Android apps. Epicc~\cite{epicc} identifies ICC connection points by using inter-procedural data-flow and points-to analysis. IC3 ~\cite{ic3} improves Epicc by resolving targets and values used in ICC. Tsutano et al.~\cite{jitana} propose JITANA to analyze interacting apps. It works in a similar fashion as Epicc but instead of combing apps for analysis, it uses static class loader which allows it to analyze large number of interacting apps. IccTA~\cite{iccta} attempts to improve static taint analysis of Android apps in ICC by using IC3 to resolve ICC targets and by modeling the life-cycle and callback methods. DidFail~\cite{didfail} attempts to detect data leaks between activities through implicit intents. It does not consider other components and explicit intents. Grace et al.~\cite{capability} perform static analysis in stock Android apps released by different vendors, to check the presence of any information leak. Since vendors modify or introduce their own apps, they might also introduce new vulnerabilities. The work, however, is limited to stock apps on specific vendor devices.
 
TaintDroid~\cite{Enck2010taintDroid} is a tool for performing dynamic taint analysis. It relies on a modified Android installation that tracks tainted data at run-time. The implementation showed minimal size and computational overhead, and was effective in analyzing many real Android apps. A complementary approach is based on static analysis~\cite{Mann2012privacyLeaks}, where a type system is implemented to track security levels. It detects violations when privileged information could potentially leave the app through a sink.

Similar to our approach, the above-mentioned approaches apply static analysis techniques on mobile code to detect information flows from sources to sinks. However, our approach does not report all sensitive data flows into sinks as information leak problems because they might be intended behaviors of the app. Our approach classifies information flows in an app as {\em anomalous} only when they deviate from normal behaviors of other similar apps. In addition, it detects not only cases of information leaks, but also cases of anomalies in data flows that might reveal security defects, such as permission re-delegation vulnerabilities. 

Other closely related work is about detecting permission re-delegation vulnerabilities in apps. Felt et al.~\cite{Porter2011re-delegation} presented the permission re-delegation problems, and their approach detects them whenever there exists a path from a public entry point to a privileged API call. 
Chin et al.~\cite{Chin2011interapplication} and Lu et al.~\cite{chex} also detect permission re-delegation vulnerabilities.
However, as acknowledged by Felt et al. and Chin et al., their approaches cannot differentiate between legitimate and illegitimate permission re-delegation behaviors. 

Zhang et al.~\cite{appsealer} proposed Appsealer, a runtime patch to mitigate permission re-delegation problem. They perform static data flow analysis to determine sensitive data flows from sources to sinks and apply a patch before the invocations of privileged APIs such that the app alerts the user of potential permission re-delegation attacks and requests the user's authorization to continue. This is an alternative way of distinguishing \normal behaviors and abnormal ones by relying on the user. Lee et al.~\cite{sealant} also proposed a similar approach but they extended the Android framework to track ICC vulnerabilities instead of patching the app. Instead of relying on the user, who might not be aware of security implications, we resort to a model that reflects normal information flow behaviors to detect anomalies in the flow of sensitive information.  
\section{Anomalous Information Flow Detection}
\label{sec:model}

\subsection{Overview}

\begin{figure*}[tbh]
\centering
\includegraphics[width=0.6\textwidth]{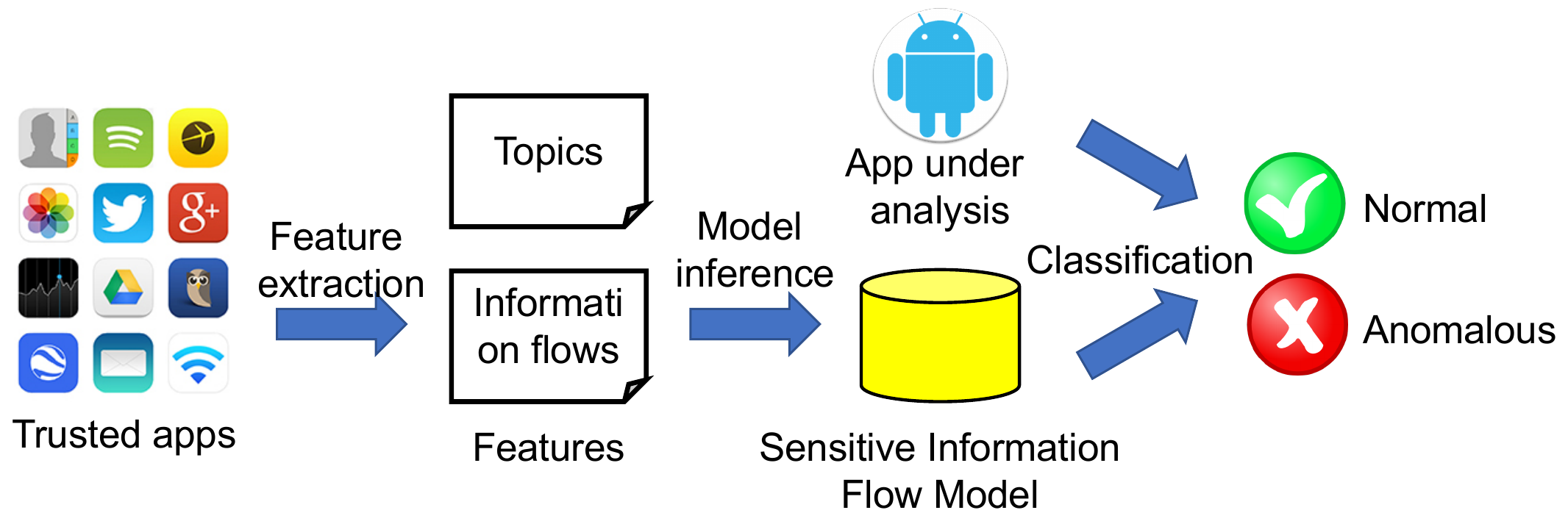}
\caption{Overview of the approach.}
\label{fig:architecture}
\end{figure*}

The overview of our approach is shown in Figure~\ref{fig:architecture}.
It has two main phases --- \emph{Learning} and \emph{Classification}. The input to the \emph{learning phase} is a set of apps that are trusted to be benign and correct in the way sensitive data is handled (we shall denote them as \emph{\reference apps}). It has two sub-steps --- feature extraction and model inference. In the feature extraction step, (i) topics that best characterize the trusted apps are inferred using natural language processing (NLP) techniques and (ii) information flows from sensitive sources to sinks in the trusted apps are identified using static taint analysis.
In the model inference step, we build sensitive information model that characterizes information flows regarding each topic.

These models and a given app under analysis (we shall denote it as \emph{\AUT}) are the inputs to the \emph{classification phase}. In this phase, basically, the dominant topic of the \AUT is first identified to determine the relevant sensitive information flow model. Then, if the \AUT contains 
any information flow that violates that model, i.e., is not consistent with the common flows characterized by the model, it is flagged as \emph{anomalous}. Otherwise, it is flagged as \emph{\normal}. 

We implemented this approach in our tool~\tool to automate the detection of anomalous information flows. However, a security analyst is required to further inspect those anomalous flows and determine whether or not the flows could actually lead to serious vulnerabilities such as information leakage issues.

\subsection{Feature Extraction}

\subsubsection{Topic Analysis}
\label{sec:topic-analysis}

In this step, we analyze the functionalities declared by the \reference apps. Using NLP, we extract topics from the app descriptions available at the official App Store, where apps developer declare their intended functionalities. 
We fist apply \emph{data pre-processing} to cleanse app descriptions and then we perform \emph{topic discovery} on the pre-processed descriptions to extract the topics that are likely to be associated with.

\emph{App descriptions pre-processing.} To apply NLP on app descriptions, we need one common language across all apps. Therefore, firstly apps with no English description are filtered. We use Google's Compact Language Detector\footnote{https://github.com/CLD2Owners/cld2} to detect English language in app descriptions.
	
Next, \emph{stopwords} in app descriptions are removed. They are words that do not contribute to topic discovery, such as  
``a'', ``after'', ``is'', ``in'', ``as'', ``very'', etc\footnote{see the list of common English stopwords at \url{www.ranks.nl/stopwords}}.  
Subsequently, the Stanford CoreNLP lemmatizer\footnote{http://stanfordnlp.github.io/CoreNLP/} is used to apply lemmatization,
which basically abstracts the words having similar meanings so that they can be analyzed as a single item.
For instance, the words ``car'', ``truck'', ``motorcycle'' appearing in the descriptions can be lemmatized as ``vehicle''. 
Last, stemming~\cite{Porter:1997:ASS:275537.275705} is applied 
to reduce words to their radix. Different forms of a word such as ``travel'', ``traveling'', ``travels'', and ``traveler'' are replaced by their common base form ``travel''.

\emph{Topics discovery.} Topics representative of a given pre-processed app description are identified using the Latent Dirichlet Allocation (LDA) technique~\cite{LDA:2003}, implemented in a tool called Mallet~\cite{Mallet:2002}. LDA is a generative statistical model that represents a collection of text as a mixture of topics with certain probabilities, where each word appearing in the text is attributable to one of the topics. The output of LDA is a list of topics, each of them with its corresponding probability. The topic with the highest probability is labeled as the \emph{dominant topic} for its associated app.

To illustrate, Figure~\ref{fig:topics} shows the functional description of an app called {\em BestTravel}, and the resulting output after performing pre-processing and topics discovery on the description. ``Travel'' is the dominant topic, the one with the highest probability of 70\%. Then, the topics ``Communication'', ``Finance'', and ``Photography'' have the 15\%, 10\%, and 5\% probabilities, respectively, of being the functionalities that the app declares to provide.

\begin{figure}[htb]
\centering
\small
\begin{tabular}{|p{0.8\linewidth}|}
\hline		
	The ultimate and most convenient way of traveling. Use BestTravel while on the move, to find restaurants (including pictures and prices), local transportation schedule, ATM machines and much more. 
 \\
\hline
\end{tabular}
\\
\vspace{2mm}
\begin{tabular}{r|cccc}		  
	App name  & Travel & Communication & Finance & Photography \\
	\hline
	BestTravel &  70\% & 15\% & 10\% & 5\% \\
\end{tabular}
\caption{Example of app description and topic analysis result.}
\label{fig:topics}
\end{figure}

Note that we did not consider Google Play categories as topics even though apps are grouped under those categories in Google Play. This is because recent studies~\cite{mharman_clustering,gorla2014checking} have reported that NLP-based topic analysis on app descriptions produces more cohesive clusters of apps than those apps grouped under Google Play categories.

\subsubsection{Static Analysis}
\label{sec:static-analysis}

Sensitive information flows in the \reference apps are extracted using static taint analysis. Taint analysis is an instance of flow-analysis technique, which
tags program data with labels that describe its provenance and propagates these tags through control and data dependencies. A different label is used for each distinct source of data. Tags are propagated from the operand(s) in the right-hand side of an assignment (uses) to the variable assigned in the left-hand side of the assignment (definition).
The output of taint analysis is information flows, i.e., what data of which provenances (\emph{sources}) are accessed at what program operations, e.g., on channels that may leak sensitive information (\emph{sinks}). 

Our analysis focuses on the flows of sensitive information into sensitive program operations, i.e., our taint analysis generates tags at API calls that read sensitive information (e.g. GPS and phone contacts) and traces the propagation of tags into API calls that perform sensitive operations such as sending messages and Bluetooth packets.
These sensitive APIs usually belong to dangerous permission group and hence, the APIs that we analyze are those privileged APIs that require to be specifically granted by the end user.
Sources and sinks are the privileged APIs available from PScout~\cite{au2012pscout}. The APIs that we analyze also include those APIs that enable Inter Process Communication (IPC) mechanism of Android because they can be used to exchange data among apps installed on the same device. 

As a result, our taint analysis generates a list of $(source\rightarrow sink)$ pairs, where each pair represents the flow of sensitive data originating from a source into a sink.

APIs (both for sources and for sinks) are grouped according to the special {\em permission} required to run them. For example, all the network related sink functions, such as {\tt openConnection()}, {\tt connect()} and {\tt getContent()} are all modeled as {\em Internet} sinks, because they all require the {\tt INTERNET} permission to be executed. 
Figure~\ref{fig:static-analysis-results} shows the static taint analysis result on the ``BestTravel'' running example app from Figure~\ref{fig:topics}. It generates two $(source\rightarrow sink)$ pairs that correspond to two sensitive information flows. In the first flow, data read from the GPS is propagated through the program until it reaches a statement where it is sent over the network. In the second flow, data from the phone contacts is used to compose a text message.

\begin{figure}[htb]
\centering
\begin{tabular}{|lcl|}
\hline
\multicolumn{3}{|l|}{App: BestTravel} \\
\hline
GPS & $\rightarrow$ & Internet \\
Contacts & $\rightarrow$ & SMS \\
\hline
\end{tabular}
\caption{Example of static analysis result.}
\label{fig:static-analysis-results}
\end{figure}

Our tool, \tool, runs on compiled byte-code of apps to perform the above static taint analysis.
It relies on two existing tools --- IC3~\cite{ic3} and IccTA~\cite{iccta}. Android apps are usually composed of several components. Therefore, to precisely extract inter-component information flows, we need to analyze the links among components.~\tool uses IC3 to resolve the target components when a flow is inter-component. IC3 uses a solver to infer all possible values of complex objects in an inter-procedural, flow- and context-sensitive manner. Once inter-component links are inferred, \tool uses an inter-component data-flow analysis tool called IccTA to perform static taint analysis. We customized IccTA to produce flows in a format as presented in Figure~\ref{fig:static-analysis-results} and paths in a more verbose format to facilitate manual checks.

\subsection{Model Inference}
\label{sec:modelInference}

When results of topic analysis and of static analysis are available for all the \reference apps, they are used to build the \emph{Sensitive Information Flow Model}. Such a model is a matrix with sensitive information sources in its rows and sinks in its columns, as shown in Figure~\ref{fig:model}.

Firstly, apps with the same dominant topic are grouped together\footnote{We  also experimented with a more elaborated model that considers multiple topics and their probabilities instead of just the dominant topic for grouping the apps. However, since their detection \ACCUN did not improve, we opted for the simplest model, with just the dominant topic.}, to build a sensitive information flow model corresponding to that specific topic. Each group is labeled with the dominant topic.
Next, each cell of the matrix is filled with a number, representing the number of apps in this group having the corresponding $(source\rightarrow sink)$ pair. 

Figure~\ref{fig:model} shows a sample sensitive information model regarding the topic ``Travel''. There are 36 distinct flows in the apps grouped under this dominant topic. The matrix shows that there are ten apps containing \emph{GPS position} flowing through the Internet (one of them being the {\em BestTravel} app, see Figure~\ref{fig:static-analysis-results}); eight apps through text messages and three apps through Bluetooth. Similarly, the matrix shows that \emph{contacts} information flows through SMS in seven apps and through Bluetooth in eight apps.

\begin{figure}[htb]
\centering
\begin{tabular}{c|ccc}
Topic: ``Travel''&\multicolumn{3}{c}{\bf Sinks} \\
\hline
{\bf Sources}& Internet & SMS & Bluetooth \\ \hline
GPS      & 10 & 8 & 3 \\
Contacts & 0  & 7 & 8 \\
\end{tabular}
\caption{Example of sensitive information flow model.}
\label{fig:model}
\end{figure}

From this model, we can observe that for Travel apps it is quite {\em common} to share the user's position via Internet and SMS. However, it is quite {\em uncommon} to share the position data via Bluetooth since it happened only in three cases. Likewise, the phone contacts are {\em commonly} shared through text messages and Bluetooth but not through Internet.

To provide a formal and operative definition of {\em common} and {\em uncommon} flows, we compute a threshold denoted as $\tau$. Flows that occur more than or equal to $\tau$ are considered as {\em common}; flows that never occur or that occur fewer than $\tau$ are considered as {\em uncommon} regarding this topic.  

Although our model assumes or trusts that the \emph{\reference} apps are benign and correct, it is possible that some of them may contain defects, vulnerabilities or malware. This problem is addressed by classifying those flows occurring less than the threshold $\tau$ as uncommon, i.e., our approach tolerates the presence of some anomalous flows in the reference model since these flows would still be regarded as uncommon. Hence, our approach works as long as the \emph{majority} of the \reference apps are truly trustworthy. 

To compute this threshold, we adopt the box-plot approach proposed by Laurikkala et al.~\cite{laurikkala2000informal}, considering only flows occurring in the model, i.e., we consider only values greater than zero. 
$\tau$ is computed in the same way as drawing outlier dots in boxplots. It is the lower quartile (25th percentile) minus the step, where the step is 1.5 times the difference between the upper quartile (75th percentile) and the lower quartile (25th percentile). 
It should be noted that $\tau$ is not trivially the lower quartile; otherwise 25\% of the apps would be {\em outliers} by construction. The threshold is lower, i.e., it is the lower quartile minus the step. Therefore, there is no fixed amount of outliers. Outliers could be few or many depending on the distribution of data. Outliers would only be those cases that are really different from the majority of the training data points. 

In the example regarding topic ``Travel'' in Figure~\ref{fig:model}, the threshold is computed considering only the five values that are $>0$. The value for the threshold is $\tau_{Travel}=7$. It means that GPS data sent through Internet (GPS $\rightarrow$ Internet) or text messages (GPS $\rightarrow$ SMS) are {\em common} for traveling apps. Conversely, even though there are three \reference apps which send GPS data through Bluetooth (GPS $\rightarrow$ Bluetooth), there are too few cases to be considered common, and this sensitive information flow will be considered {\em uncommon} in the model.
Likewise, phone contacts are {\em commonly} sent through text messages and Bluetooth, but it is {\em uncommon} for them to be sent through the Internet, since this {\em never} occurs in the \reference apps.

\subsection{Classification}

After the Sensitive Information Flow Models are built on \reference apps, they can be used to classify a new \AUT. First of all, features must be extracted from the \AUT. The features are the topics associated with the app description and the sensitive information flows in the app. 
As in Section~\ref{sec:topic-analysis}, first data pre-processing is performed on the app description of the \AUT. Then, topics and their probabilities are inferred from the pre-processed description using the Mallet tool. Among all the topics, we consider only the {\em dominant topic}, the one with the highest probability, because it is the topic that most characterizes this app. We then obtain the Sensitive Information Flow Model associated with this dominant topic.

 To ensure the availability of the Sensitive Information Flow Model, the Mallet tool is configured with the list of topics for which the Models are already built on the \reference apps. And given an app description, the Mallet tool only generates topics from this list. The more diverse \reference apps we analyze, the more complete list of models we expect to build.

For example, Figure~\ref{fig:aut_example}(a) shows the topics inferred from the description of a sample \AUT ``TripOrganizer''. The topic ``Travel'' is highlighted in bold to denote that it is the dominant topic. 

Next, sensitive information flows in the \AUT are extracted as described in Section~\ref{sec:static-analysis}. The extracted flows are then compared against the flows in the model associated with the dominant topic. If the \AUT contains only flows that are {\em common} according to the model, the app is considered consistent with the model. If the app contains a flow that is not present in the model or a flow that is present but is {\em uncommon} according to the model, the flow and thus, the app is classified as anomalous. 
Anomalous flows require further manual inspection by a security analyst, because they could be due to defects, vulnerabilities, or malicious intentions.

For example, Figure~\ref{fig:aut_example}(b) shows three sensitive information flows extracted from ``TripOrganizer'' app. Since the dominant topic for this app is ``Travel'', these flows can be checked against the model associated with this topic shown in Figure~\ref{fig:model}. Regarding this model, earlier, we computed that the threshold is $\tau_{Travel}=7$ and the flow (Contacts $\rightarrow$ SMS) is common (see Section~\ref{sec:modelInference}). Therefore, flow 1 observed in ``TripOrganizer'' (Figure~\ref{fig:aut_example}(b)) is consistent with the model. However, flow 2 (Contacts $\rightarrow$ Internet) and flow 3 (GPS $\rightarrow$ Bluetooth), highlighted in bold in Figure~\ref{fig:aut_example}(b), are uncommon according to the model. As a result, the \AUT ``TripOrganizer'' is classified as \emph{anomalous}.

\begin{figure}[htb]
	\centering

    \begin{tabular}{r|cccc}		  
	App name  & {\bf Travel} & Books & Tools & Game \\
	\hline
	TripOrganizer &   {\bf 78\%} & 11\% & 4\% & 7\% \\
    \end{tabular}\\
    \vspace{2mm} 
    (a) Topics classification\\
	\vspace{2mm} 
	\begin{tabular}{|llcl|}
		\hline
		\multicolumn{4}{|l|}{App: TripOrganizer} \\
		\hline
		1: & Contacts & $\rightarrow$ & SMS \\
		{\bf 2:} & {\bf Contacts} & {\bf $\rightarrow$} & {\bf Internet} \\
		{\bf 3:} & {\bf GPS} & {\bf $\rightarrow$} & {\bf Bluetooth} \\
		\hline
	\end{tabular}\\
	\vspace{2mm} 
   (b) Sensitive information flows\\
	\caption{Classification of the app under analysis.}
	\label{fig:aut_example}
\end{figure}

\section{Empirical Assessment}
\label{sec:results}

In this section, we evaluate the usefulness of our approach and  report the results. We assess our approach by answering the following research questions:

\begin{itemize}
\item {\bf RQ}$_{Vul}$: Is \tool useful for identifying {
\em vulnerable apps} containing anomalous information flows?
\item {\bf RQ}$_{Time}$: How {\em long} does \tool take to classify apps?
\item {\bf RQ}$_{Topics}$: Is the {\em topic} feature really needed to detect anomalous flows?
\item {\bf RQ}$_{Cat}$: Can app-store {\em categories} be used instead of {\em topics} to learn an \ACCU Sensitive Information Flow Model?
\item {\bf RQ}$_{Mal}$: Is \tool useful for identifying {\em malicious} apps? 
\end{itemize}

The first research question {\bf RQ}$_{Vul}$ investigates the result of \tool, whether it is useful for detecting anomalies in vulnerable apps that, for example, may leak sensitive information.
{\bf RQ}$_{Time}$ investigates the cost of using our approach in terms of the time taken to analyze a given \AUT. A short analysis time is essential for tool adoption in a real production environment. 

Then, in the next two research questions, we investigate the role of {\em topics} as a feature for building the Sensitive Information Flow Models. {\bf RQ}$_{Topics}$ investigates the absolute contribution of topics, by learning the Sensitive Information Flow Model without considering the topics and by comparing its performance with that of our original model. To answer {\bf RQ}$_{Cat}$, we replace topics with the {\em categories} defined in the official market, and we compare the performance of this new model with that of our original model. 

Finally, the last research question {\bf RQ}$_{Mal}$ investigates the usefulness of \tool in detecting malware based on anomalies in sensitive information flows.

\subsection{Benchmarks and Experimental Settings}

\subsubsection{\Reference Apps}

\tool needs a set of \reference apps to learn what is the {\em normal} behavior for ``correct and benign'' apps. We defined the following guidelines to collect \reference apps: (i) apps that come from the official Google Play Store (so they are scrutinized and checked by the store maintainer) and (ii) apps that are very popular (so they are widely used and reviewed by a large community of end users and programming mistakes are quickly notified and patched). 

At the time of crawling the Google Play Store, it had 30 different app categories. From each category, we downloaded, on average, the top \avgnumAppsPerCat apps together with their descriptions. We then discarded apps with non-English description and those with very short descriptions (less than 10 words). Eventually, we are left with \numApps apps for building references models. 

Additionally, we measured if these apps were actively maintained by looking at the date of the last update. 70\% of the apps were last updated in the past 6 months before the Play Store was crawled, while 32\% of the apps were last updated within the same month of the crawling. This supports the claim that the \reference apps are well maintained.

The fact that the \reference apps we use are suggested and endorsed by the official store, and that they collected good end-user feedback allows us to assume that the apps are of high quality and do not contain many security problems. Nevertheless, as explained in Section~\ref{sec:modelInference}, our approach is robust against the inclusion of a small number of anomalous apps in the training set since we adopt a threshold to classify anomalous information flows.

\subsubsection{Subject Benign Apps}
\label{subsec:benignapps}

\tool works on compiled apps and, therefore the availability of source code is not a requirement for the analysis. However, for this experiment sake, we opted for open source projects, which enable us to inspect the source code and establish the \emph{ground truth}. 

The F-Droid repository\footnote{\texttt{http://f-droid.org/}} represents an ideal setting for our experimentation because (i) it includes real world apps that are also popular in the Google Play Store, and (ii) apps can be downloaded with their source code for manual verification of the vulnerability reports delivered by \tool.

The F-Droid repository was crawled in July 2017 for apps that meet our criteria. Among all the apps available in this repository, we used only those apps that are also available in the Google Play Store, whose descriptions meet our selection criteria (i.e., description is in English and it is longer than 10 words). Eventually, our experimental set of benign apps consists of \numAUTTotal \AUTs.

\subsubsection{Subject Malicious Apps}
\label{subsec:malware}

To investigate if \tool can identify malware, we need a set of malicious apps with their \emph{declared} functional descriptions. 
Malicious apps are usually repackaged versions of popular (benign) apps, injected with malicious code (Trojanized); hence the descriptions of those popular apps they disguise as can be considered as their app descriptions. Hence, by identifying the original versions of these malicious apps in the Google Play Store, we obtain their declared functional descriptions.

We consider the malicious apps from the Drebin malware dataset~\cite{arp2014drebin}, which consists of 5,560 samples that have been collected in the period of August 2010 to October 2012. We randomly sampled 560 apps from this dataset. For each malicious app, we performed static analysis to extract the {\em package name}, an identifier used by Android and by the official store to distinguish Android apps\footnote{Even if it is easy to obfuscate this piece of information, in our experiment some apps did not rename their package name}. We queried the official Google Play market for the original apps, by searching for those having the same package name. Among our sampled repackaged malicious apps, we found 20 of the apps in the official market with the same package name. We analyzed their descriptions and found that only 18 of them have English descriptions. We therefore performed static taint analysis on these 18 malware samples, for which we found their ``host'' apps in the official market. Our static analysis crashed on 6 cases. Therefore, our experimental set of malicious apps consists of 12 \AUTs.

\subsection{Results}

\subsubsection{Detecting Vulnerable Apps}
\label{subsec:analysis1}
\begin{table*}[htb]
\caption{Anomaly detection based on Sensitive Information Flows Model} 
\centering
\small
\begin{tabular}{llll}
  \hline
App & Topic & Source & Sink \\ 
  \hline
\multirow{5}{*}{a2dp.Vol}& \multirow{5}{*}{communications utility} & Bluetooth & IPC \\ 
   &  & Bluetooth & Modify audio settings\\ 
   &  & Bluetooth admin & Modify audio settingss \\ 
   &  & Broadcast sticky & Modify audio settingss \\ 
   &  & Modify audio settings & Modify audio settingss \\ 
 \cmidrule{1-4}      
  
  com.alfray.timeriffic & tools and utility & IPC & Write settingss \\ 
  \cmidrule{1-4}
  com.dozingcatsoftware.asciicam & photo editors & Camera & IPCs \\ 
  \cmidrule{1-4}  
  
  \bf{com.matoski.adbm} & \bf{utility} & \bf{Access WiFi state} & \bf{IPC} \\ 
  \cmidrule{1-4}

  \bf{com.mschlauch.comfortreader} & \bf{books \& readers} & \bf{IPC} & \bf{Internet} \\ 
\cmidrule{1-4}

  com.newsblur & productivity & IPC & Access network states \\ 
    \cmidrule{1-4}   
\multirow{5}{*}{com.Pau.ImapNotes2} & \multirow{5}{*}{documents manager}& Authenticate accounts & IPCs \\ 
 &  & Authenticate accounts & Authenticate accountss \\ 
  &  & Get accounts & Authenticate accountss \\ 
  &  & Get accounts & Get accountss \\ 
  &  & Get accounts & Manage accountss \\ 
   \cmidrule{1-4}
  \multirow{2}{*}{com.thibaudperso.sonyCamera }  &   \multirow{2}{*}{tools \& utility} & IPC & Change WiFi states \\ 
  & & NFC & NFCs \\
  \cmidrule{1-4}     
    fr.neamar.kiss & security tools & IPC & Access WiFi states \\  
\cmidrule{1-4}

  \multirow{2}{*}{fr.ybo.transportsrennes} &  \multirow{2}{*}{utility} & Access coarse location & Internets \\ 
  &  & Access fine location & Internets \\ 
 \cmidrule{1-4}        

 \multirow{2}{*}{mobi.boilr.boilr}  & \multirow{2}{*}{financial}  & IPC & Wake lock  \\ 
   &  & Wake lock & Wake lock  \\ 
  \cmidrule{1-4}        

  org.smc.inputmethod.indic & languages learning & IPC & Vibrate  \\ 
\cmidrule{1-4}

  pw.thedrhax.mosmetro & books \& readers & Access WiFi state & Change WiFi state  \\ 
\cmidrule{1-4}

  se.anyro.Nfc\_reader & communications utility & IPC & NFC  \\ 
   \hline
\end{tabular}
\label{tab:modelResult}
\end{table*}

Firstly, \tool was used to perform static taint analysis on the \numApps \reference apps and topic analysis on their descriptions from the official Play Store. It then learns the Sensitive Information Flow Models based on the dominant topics and extracted flows as described in Section~\ref{sec:modelInference}. Then, the \AUTs from the F-Droid repository (Section~\ref{subsec:benignapps}) have been classified based on the Sensitive Information Flow Models.

Out of \numAUTTotal \AUTs, static taint analysis reported \numAUTWithRelevantFlows apps to contain flows of sensitive information that reach sinks, for a total of 1428 flows. These flows map to 147 distinct source-sink pairs. Out of these \numAUTWithRelevantFlows apps, 14 \AUTs are classified as {\em anomalous}. Table~\ref{tab:modelResult} shows the analysis results reported by \tool. The first column presents the name of the app. The second column presents the app's dominant topic. The third and fourth columns present the source of sensitive data and the sink identified by static taint analysis, respectively. 
As shown in Table~\ref{tab:modelResult}, in total \tool reported \numFlows anomalous flows in these apps.
We manually inspected the source code available from the repository to determine if these anomalous flows were due to programming defects or vulnerabilities. Two apps are found to be vulnerable (highlighted in boldface in Table~\ref{tab:modelResult}), they are {\em com.matoski.\newline adbm} and {\em com.mschlauch.comfortreader}. 

{\em com.matoski.adbm} is a utility app for managing the ADB debugging interface. 
The anomalous flow involves data from the WiFi configuration that leak to other apps through the Inter Process Communication. Among other information that may leak, the SSID data, which identifies the network to which the device is connected to, can be used to infer the user position and threaten the end user privacy. Hence, this programming defect leads to information leakage vulnerability that requires corrective maintenance. We reported this vulnerability to the app owners on their issue tracker. 

{\em com.mschlauch.comfortreader} is a book reader app, with an anomalous flow of data from IPC to the Internet. Manual inspection revealed that this anomalous flow results from a permission re-delegation vulnerability because data coming from another app is used, without sanitization, for opening a data stream. If a malicious app that does not have the permission to use the Internet passes a URL that contains sensitive privacy data (e.g., GPS coordinates), then the app could be used to leak information. We reported this vulnerability to the app developers. 

Regarding the other 12 \AUTs, even though they contain anomalous flows compared to \reference apps, manual inspection revealed that they are neither defective nor vulnerable. For example, some apps contain anomalous flows that involves IPC. Since data may come from other apps via IPC (source) or may flow to other apps via IPC (sink), such flows are considered dangerous in general. However, in these 12 apps, when IPC is a source (e.g., in {\tt com.alfray.timeriffic}), data is either validated/sanitized before used in the sink or used in a way that do not threaten security. 
On the other hand, when IPC is a sink (e.g., in {\tt com.dozingcatsoftware.\newline asciicam}), the destination is always a component in the same app, so the flows are not actually dangerous.

Since \tool helped us detect 2 vulnerable apps containing anomalous information flows, 
we can answer {\bf RQ}$_{Vul}$ by stating that \tool is useful for identifying vulnerabilities related to anomalous information flows.

\subsubsection{Classification Time}
To investigate {\bf RQ}$_{Time}$, we analyze the time required to classify the \AUTs.
We instrumented the analysis script with the Linux {\em date} utility to log the time (in seconds) before starting the analysis and at its conclusion. Their difference is the amount of time spent in the computation. The experiment was run on a multi-core cluster, specifically designed to let a process run without sharing memory or computing resources with other processes. Thus, we assume that the time measurement is reliable.

\begin{figure}[htb]
\centering
\includegraphics[width=0.3\textwidth,trim={0 2.5cm 0 2cm},clip]{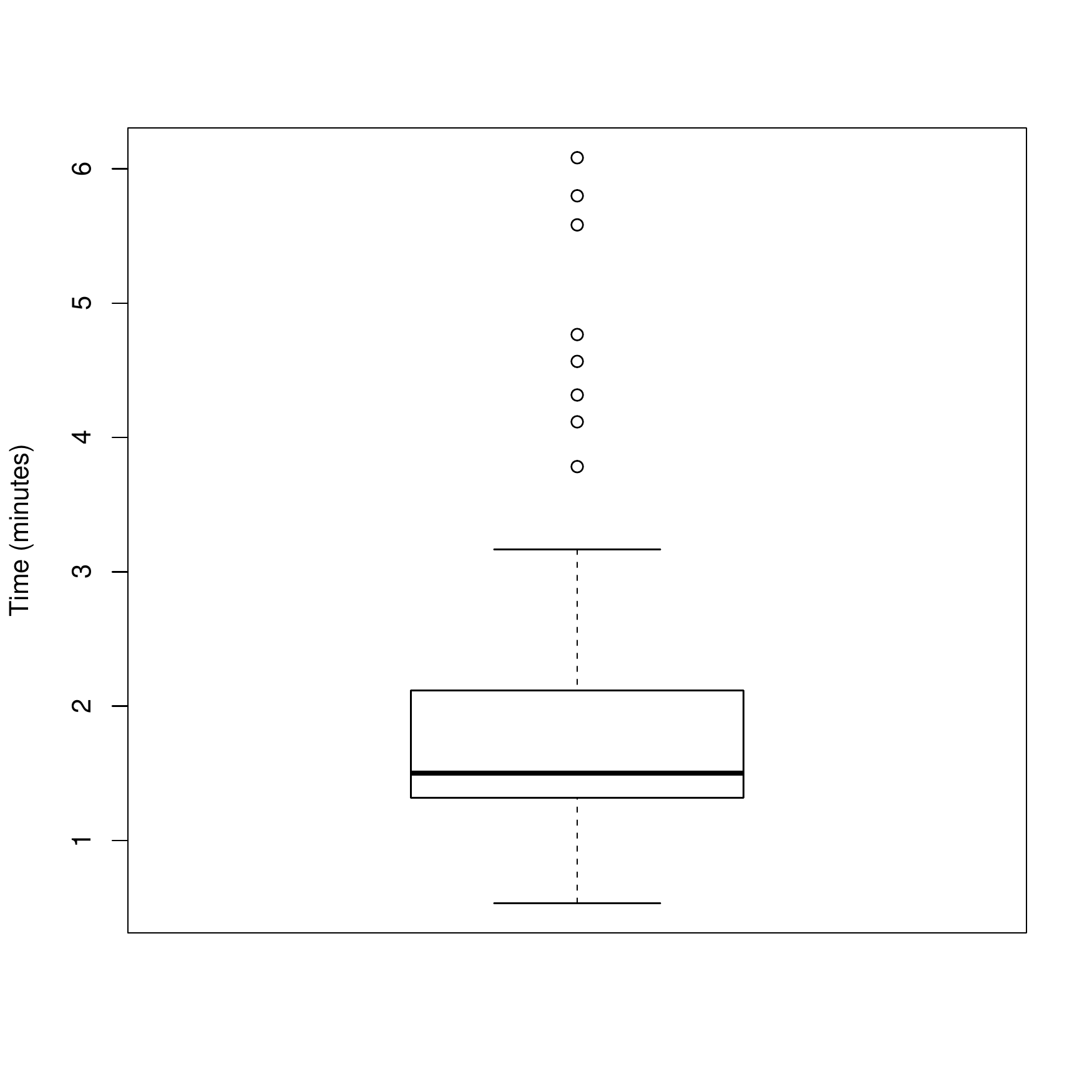}
\begin{tabular}{ccc}
  \hline
 Mean & Median & Standard Deviation \\ 
  \hline
1.9 & 1.5 & 1.07 \\ 
   \hline
\end{tabular}
\caption{Boxplot of classification time.}
\label{fig:time}
\end{figure}

Classification time includes the static analysis step to exact data flow, the natural language step to extract topics from description and the comparison with the Sensitive Information Flow Model to check for consistency. Figure~\ref{fig:time} reports the boxplot of the time (in minutes) needed to classify the F-Droid apps and the descriptive statistics. On average, an app takes 1.9 minutes to complete the classification and most of the analyses concluded in less than 3 minutes (median = 1.5). Only a few (outliers) cases require longer analysis time.

\subsubsection{Topics from App Description}

\begin{table*}[htb]
\caption{Anomaly detection using only information flows as a feature (no topic feature)}
\centering
\small
\begin{tabular}{llll}
  \hline
App & Topic & Source & Sink  \\ 
  \hline
\multirow{4}{*}{a2dp.Vol} & \multirow{4}{*}{Communications utility} & Bluetooth & Modify audio settings  \\ 
  &  & Bluetooth admin & Modify audio settings  \\ 
  &  & Broadcast sticky & Modify audio settings  \\ 
  &  & Modify audio settings & Modify audio settings  \\ 
  \cmidrule{1-4}
  \multirow{4}{*}{com.Pau.ImapNotes2} & \multirow{4}{*}{Document manager} & Authenticate accounts & IPC  \\ 
   & & Authenticate accounts & Authenticate accounts  \\ 
    & & Get accounts & Authenticate accounts  \\ 
   & & Get accounts & Get accounts  \\ 
   & & Get accounts & Manage accounts  \\ 
   \cmidrule{1-4}
  \multirow{2}{*}{com.thibaudperso.sonyCamera} & \multirow{2}{*}{Utility} & IPC & Change wifi state  \\ 
   & & NFC & NFC  \\ 
   \cmidrule{1-4}
  se.anyro.Nfc\_reader & Communications utility & IPC & NFC  \\ 
   \hline
\end{tabular}
\label{tab:noTopicResults} 
\end{table*}

We now run another experiment to verify our claim that topics are important features to build an \ACCU model \newline({\bf RQ}$_{Topics}$). We repeated the same experiment as before, but using only flows as features and without considering topics, to check how much detection \ACCUN we lose in this way.

We still consider all the \reference apps for learning the reference model, but we only use static analysis data. That is, we do not create a separate matrix for each topic; instead we create one big single matrix with sources and sinks for all the apps. This Sensitive Information Flow Model is then used to classify F-Droid apps and the results are shown in Table~\ref{tab:noTopicResults}. As we can see, only four apps are detected as {\em anomalous} by this second approach, and all of them were already detected by our original, proposed approach. Manual inspection revealed that all of them are not vulnerable.

This suggests that topic is a very important feature to learn reference models in order to detect a larger amount of anomalous apps. 
In fact, when topics are not considered and all the apps are grouped together regardless of their topics, we observe a {\em smoothing} effect. Differences among apps become less relevant to detect anomalies. While in the previous model, an app was compared only against those apps grouped under the same topic. Here, an app is compared to all the \reference apps. Without topic as a feature, our model loses the ability to capture the characteristics of distinct groups and, thus, the ability to detect deviations from them.

\subsubsection{Play Store Categories}

\begin{table*}[htb]
\caption{Anomaly detection using Google Play categories as a feature instead of topics} 
\centering
\small
\begin{tabular}{llll}
  \hline
App & Category & Source & Sink  \\ 
  \hline
\multirow{6}{*}{a2dp.Vol} & \multirow{6}{*}{Transportation} & Bluetooth & IPC  \\ 
  &  & Bluetooth & Bluetooth  \\ 
  &  & Bluetooth & Modify audio settings  \\ 
  &  & Bluetooth admin & Modify audio settings  \\ 
   &  & Broadcast sticky & Modify audio settings  \\ 
   &  & Modify audio settings & Modify audio settings  \\
    \cmidrule{1-4}   
  \multirow{5}{*}{com.Pau.ImapNotes2} & \multirow{5}{*}{Productivity} & Authenticate accounts & IPC  \\ 
  &  & Authenticate accounts & Authenticate accounts  \\ 
   & & Get accounts & Authenticate accounts  \\ 
  &  & Get accounts & Get accounts  \\ 
  &  & Get accounts & Manage accounts  \\ 
     \cmidrule{1-4}
  
  com.alfray.timeriffic & Tools & IPC & Write settings  \\ 
  \cmidrule{1-4}

  com.angrydoughnuts.android.alarmclock & Tools & Wake lock & Vibrate  \\ 
  \cmidrule{1-4}

  com.dozingcatsoftware.asciicam & Photography & Camera & IPC  \\ 
  \cmidrule{1-4}
  
  com.futurice.android.reservator & Business & Get accounts & IPC  \\ 
 \cmidrule{1-4}
 
  \bf{com.matoski.adbm} & \bf{Tools} & \bf{Access wifi state} & \bf{IPC}  \\
      \cmidrule{1-4}

  \multirow{2}{*}{com.thibaudperso.sonyCamera}& \multirow{2}{*}{Media and video} & IPC & Change wifi state  \\ 
   & & NFC & NFC  \\ 
  \cmidrule{1-4}
  mobi.boilr.boilr & Finance & Wake lock & Wake lock  \\ 

  \cmidrule{1-4}
  se.anyro.Nfc\_reader & Communication & IPC & NFC  \\ 
   \hline
\end{tabular}
\label{tab:resultsGoogleCategories}
\end{table*}

To investigate {\bf RQ}$_{Cat}$, instead of grouping \reference apps based on topics, we group them according to their app categories as determined by the official Google Play Store. First of all we split \reference apps into groups based on the market category they belong to\footnote{At the time of the crawling, in Google Play Store, a popular app was assigned to only one category.}. We then use static analysis information about flows to build a separate source-sink matrix per each category. Eventually we compute thresholds to complete the model.

We then classify each \AUT from F-Droid by comparing it with the model of the corresponding market category. The classification results are reported in Table~\ref{tab:resultsGoogleCategories}. Ten apps are reported as containing anomalous flows and most of them were also detected by our original, proposed approach (Table~\ref{tab:modelResult}). Two apps reported by this approach were not reported by our proposed approach, which are {\em com.angrydoughnuts.android.alarmclock} and {\em com.futurice.\newline android.reservator}. However, they are neither the cases of vulnerabilities nor malicious behaviors. Only one flow detected by this approach is a case of vulnerability, namely {\em com.matoski.adbm}, highlighted in boldface, which was also detected by our proposed approach. 
Hence, this result supports our design decision of using topics.

\subsubsection{Comparison of the Models}

\begin{table}[htb]
\caption{Summary of model comparison result} 
\centering
\begin{tabular}{l|l|rrr}
  \hline
RQ & Features & FP & Unique & Vuln. \\ 
  \hline
RQ$_{Vul}$ & Flows + Topics      		& 12 & 5 & 2 \\
RQ$_{Topics}$ & Flows         			& 4 & 0 & 0 \\
RQ$_{Cat}$ & Flows + Market Cat. 	& 9 & 2 & 1 \\
   \hline
\end{tabular}
\label{tab:summary}
\end{table}

Table~\ref{tab:summary} summarizes the result of the models comparison. The first model (first row) considers both data flows and description topics as features. Even though this approach reported the largest number of false positives (12 apps, `FP' column), we were able to detect 2 vulnerabilities (`Vuln.' column) by tracing the anomalies reported by this approach. 
It also detected 5 additional anomalous apps that other approaches did not detect (`Unique' column).

The second model (second row) considers only data flows as a feature. Even though the number of false positives drops to 4, we were not able to detect any vulnerability by tracing the anomalies reported by this approach. This result suggests that modeling only flows is not enough for detecting vulnerabilities. 

When market categories are used instead of description topics (last row), the false positives drops to 9 (25\% less compared to our proposed model). It detected 2 additional anomalous apps that other approaches did not detect (`Unique' column). Tracing the anomalies reported by this approach, we detected only one out of the two vulnerabilities that we detected using our proposed approach. This result suggests that topics are more useful than categories for detecting vulnerable apps containing anomalous information flows.

\subsubsection{Detecting Malicious Apps}
\label{subsec:malwdetect}
Anomalies in the flow of sensitive data could be due to malicious behaviors as well. The goal of this last experiment is to investigate whether \tool can be used to identify malware ({\bf RQ}$_{Mal}$). To this aim, we use the Sensitive Information Flow Model (learned on the \reference apps) to classify the 18 \AUTs from the Drebin malware dataset. Data flow features are extracted using static analysis from these malicious apps. However, static taint analysis crashed on 6 apps because of their heavy obfuscation. Since improving the static taint analysis implementation to work on heavy obfuscated code is out of the scope of this paper, we run the experiment on the remaining 12 apps.  Topics are extracted from the descriptions of the original versions of those malware, which are available at the official market store.

\begin{table*}[ht]
\caption{Results on malicious apps}
	\centering
	\tiny
	\begin{tabular}{ll|ccc}
		\hline
		Malware Name & Repackage App Name & Flows + Topics & Flows only & Flows + Market Cat. \\ 
		\hline
		PJApps & com.appspot.swisscodemonkeys.steam &\ding{52} &  \ding{52}  & \ding{52} \\ 
		DroidKungFu (variant 1) & com.gp.jaro (variant) & \ding{52} & \ding{52}   & \ding{52} \\ 	
		DroidKungFu (variant 2) & com.gp.jaro (variant) & \ding{52} &  \ding{54}  &  \ding{52} \\ 
		Spy.GoldDream& com.rechild.advancedtaskkiller & \ding{52} &  \ding{52}  &  \ding{52}   \\ 		
		Anserver & com.sohu.blog.lzn1007.WatermelonProber & \ding{52} &  \ding{54}  &  \ding{52}   \\ 
		TrojanSMS.Agent & org.baole.app.blacklistpro & \ding{52} &  \ding{52}  &  \ding{52} \\ 
		\hline
	\end{tabular}
	\label{fig:malwaredata}
\end{table*}

The malicious apps have been subject to anomaly detection, based on the three distinct feature sets: (i) flows and topics; (ii) only flows; and (iii) flows and market categories. The classification results are shown in Table~\ref{fig:malwaredata}. The first column reports the malware name (according to ESET-NOD32\footnote{https://www.eset.com/} antivirus) and the second column contains the name of the original app that was repackaged to spread the malware. The remaining three columns report the results of malware detection by the three models based on different sets of features: a tick mark (``\ding{52}'') means that the model correctly detected the app as anomalous, while a cross (``\ding{54}'') means no anomaly detected.

While the model based on topics and the model based on market categories classified the same 6 \AUTs as malicious, the model based on only flows classified only 4 \AUTs as malicious.

All the malware except {\tt TrojanSMS.Agent} are the cases of privacy sensitive information leaks such as device ID, phone number, e-mail or GPS coordinate, being sent over the network or via SMS. 
One typical malicious behavior is observed in {\tt Spy.GoldDream}. In this case, after querying the list of installed packages (sensitive data source), the malware attempts to kill selected background processes (sensitive sink). This is a typical malicious behavior observed in malware that tries to avoid detection by stopping security products such as antiviruses.
Botnet behavior is observed in {\tt DroidKunFu}. A command and control (C\&C) server command is consulted (sensitive source) before performing privileged actions on the device (sensitive sink).

As shown in Table~\ref{fig:malwaredata}, when only static analysis features are used in the model, two malicious apps are missed. This is because this limited model compares the given \AUT against all the \reference apps, instead of only the apps from a specific subset (grouped by the common topic or the same category). A flow that would have been anomalous for the specific topic (or the specific category) might be \normal for another topic/category. For example, acquiring GPS coordinate and sending it over the network is common for navigation or transportation apps. However, it is not a common behavior for tools apps, which is the case of the {\tt Anserver} malware.

The remaining 6 apps in the dataset were consistently classified as not-anomalous by all the models. These false negatives are mainly due to the malicious behaviors not related to sensitive information flows, such as dialing calls in the background or blocking messages. Another reason is due to the obfuscation by malware to hide the sensitive information flows. Static analysis inherently cannot handle obfuscation.

\subsection{Limitation and Discussion}

In the following, we discuss some of the limitations of our approach and of its experimental validation. 
The most prominent limitation to adopt our approach is the availability of \reference apps to build the model of sensitive information flows. In our experimental validation, we trusted top ranked popular apps from the official app store, but we have no guarantee that they are all immune from vulnerabilities and from malware content. However, as explained in Section~\ref{sec:modelInference}, our approach is quite robust with respect to the inclusion of a small number of defective, vulnerable, or malicious apps in the training set, as long as the majority of the training apps are benign and correct. This is because we use a threshold-based approach that models flows common to a large set of apps. Thus, vulnerable flows occurring on few training apps are not learnt as \normal in the model and they would be classified as {\em anomalous} when observed in a given \AUT.

A flow classified as {\em anomalous} by our model needs further manual analysis to check if the anomaly is a vulnerability, a malicious behavior or is safe. Manual inspection could be an expensive task that might delay the delivery of the software product. However, in our experimental validation, manual filtering on the experimental result took quite short time, on average 30 minutes per app. Considering that the code of the app to review was new to us, we expect a shorter manual filtering phase for a developer who is quite familiar with the code of her/his app. All in all, manual effort required to manual filter results of the automated tool seems to be compatible with the fast time-to-market pressure of smart phone apps.

When building sensitive information flow models, we also considered grouping of apps by using clustering technique based on the topics distribution, instead of grouping based on the dominant topic alone. But we conducted  preliminary experiments using this method and observed that grouping of apps based on dominant topics produce more cohesive groups, i.e., apps that are more similar.

Inherently, it is difficult for static analysis-based approaches including ours to handle obfuscated code. Therefore, if training apps are obfuscated (e.g., to limit reverse engineering attacks), our approach may collect incomplete static information and only build a partial model. And if the \AUT is obfuscated,  our approach may not detect the anomalies.
As future work, we plan to incorporate our approach with dynamic analysis to deal with obfuscation.

 \section{Conclusion}
\label{sec:conclusion}

In this paper, we proposed a novel approach to analyze the flows of sensitive information in Android apps. In our approach, \reference apps are first analyzed to extract topics from their descriptions and data flows from their code. Topics and flows are then used to learn Sensitive Information Flow models. We can use these models for analyzing new Android apps to determine whether they contain anomalous information flows. Our experiments show that this approach could detect anomalous flows in vulnerable and malicious apps quite fast. 
\section*{Acknowledgment}
This work has partially been supported by the activity ``API Assistant'' of the action line Digital Infrastructure of the EIT Digital and the GAUSS national research project, which has been funded by the MIUR under the PRIN 2015 program (Contract 2015KWREMX). 
The work of L. K. Shar has been partially supported by the National Research Fund, Luxembourg, while the author was affiliated with the University of Luxembourg (INTER/AAL/15/11213850 and INTER/DFG/14/11092585).

\bibliographystyle{abbrv}
\bibliography{bib}

\end{document}